\documentclass[aps,prl,twocolumn,nofootinbib,superscriptaddress]{revtex4-1}  
\pdfoutput=1

\usepackage{graphicx}  
\usepackage{amsmath,amssymb}  
\usepackage{hyperref}
\usepackage{color}
\graphicspath{{./Figures/}}

\hypersetup{
    colorlinks=true,    
    linkcolor=red,      
    citecolor=blue,     
    filecolor=magenta,  
    urlcolor=blue       
}

\begin{document}

\title{Pure Natural Inflation}

\author{Yasunori Nomura}
\affiliation{Berkeley Center for Theoretical Physics, Department of Physics,
University of California, Berkeley, CA 94720, USA}
\affiliation{Theoretical Physics Group, Lawrence Berkeley National Laboratory,
Berkeley, CA 94720, USA}
\affiliation{Kavli Institute for the Physics and Mathematics of the 
 Universe (WPI), University of Tokyo, 
 Kashiwa, Chiba 277-8583, Japan}

\author{Taizan Watari}
\affiliation{Kavli Institute for the Physics and Mathematics of the 
 Universe (WPI), University of Tokyo, 
 Kashiwa, Chiba 277-8583, Japan}

\author{Masahito Yamazaki}
\affiliation{Kavli Institute for the Physics and Mathematics of the 
 Universe (WPI), University of Tokyo, 
 Kashiwa, Chiba 277-8583, Japan}

\begin{abstract}
We point out that a simple inflationary model in which the axionic inflaton 
couples to a pure Yang-Mills theory may give the scalar spectral index 
($n_s$) and tensor-to-scalar ratio ($r$) in complete agreement with the 
current observational data.
\end{abstract}


\maketitle

Cosmic inflation plays an important role in explaining observable features 
of our universe, including its extreme flatness, as well as the origin 
of primordial curvature perturbations.  The detailed predictions of inflation, 
however, depend on the potential $V(\phi)$ of the inflaton field $\phi$. 
An important issue, therefore, is to understand what is the correct model 
of inflation and how it emerges from the underlying physics.

Recent observations by Planck~\cite{Ade:2015lrj} and BICEP2/Keck 
Array~\cite{Array:2015xqh} have started constraining simple models 
of inflation.  In particular, arguably the simplest model of inflation 
$V(\phi) = m^2 \phi^2/2$~\cite{Linde:1983gd}---which gives the correct 
value for the scalar spectral index $n_s \simeq 0.96$---is now excluded 
at about the $3\sigma$ level because of the non-observation of tensor 
modes.  This raises the following questions.  Does the model of inflation 
need to be significantly complicated?  Is the agreement of $n_s$ of the 
quadratic potential with the data purely accidental?

In this letter, we argue that the answers to these questions may both be 
no.  In particular, we argue that a simple inflationary model in which the 
inflaton $\phi$ couples to the gauge field of a pure Yang-Mills theory
\begin{equation}
  \mathcal{L} = \frac{1}{32\pi^2} \frac{\phi}{f}\, 
    \epsilon^{\mu\nu\rho\sigma}\, \textrm{Tr}\, F_{\mu\nu} F_{\rho\sigma},
\label{eq:L-theta}
\end{equation}
may give the values of $n_s$ and the tensor-to-scalar ratio, $r$, in 
perfect agreement with the current observational data.  Here, $\phi$ is 
a pseudo-Nambu-Goldstone boson---axion---of a shift symmetry $\phi \to 
\phi + \textrm{const.}$, and $f$ is the axion decay constant.  For now 
we assume that the gauge group of the Yang-Mills theory is $SU(N)$, 
but the model also works for other gauge groups; see later.

\begin{figure}
  \includegraphics[scale=0.68]{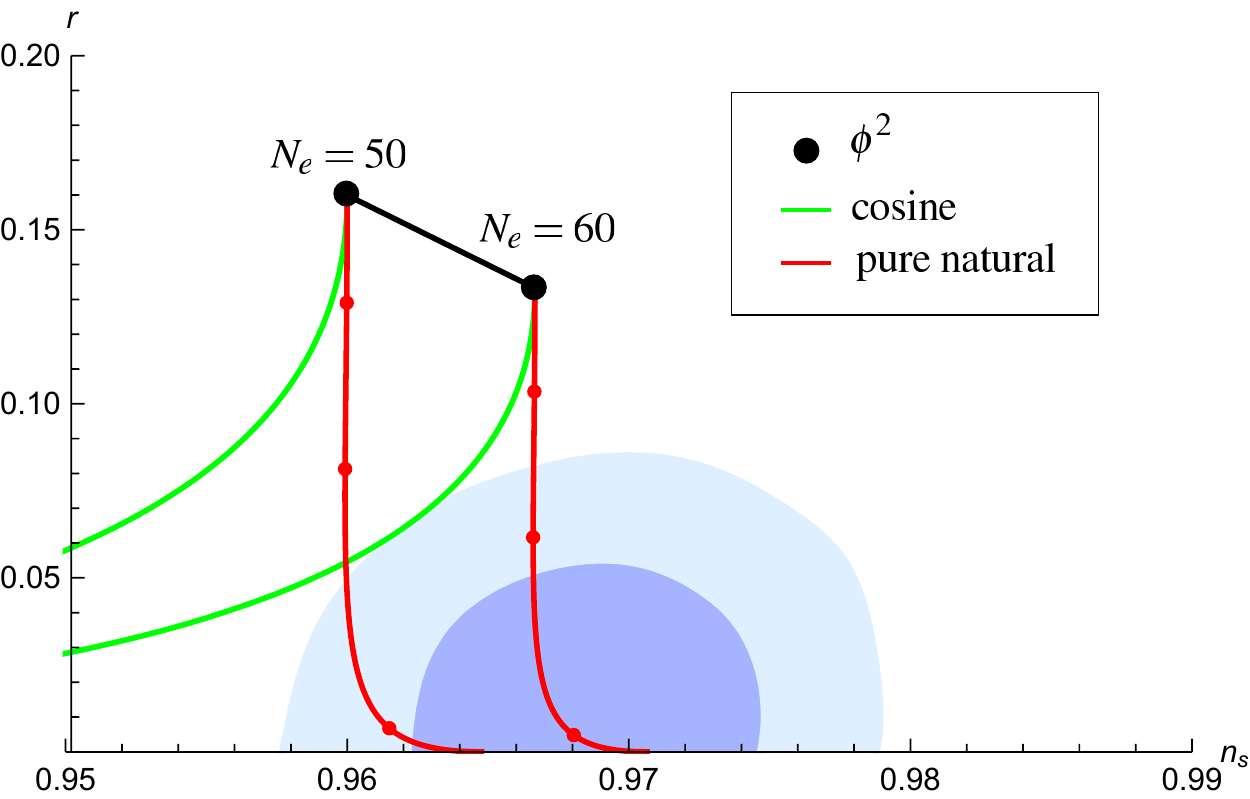}
\caption{The predicted values of $n_s$ and $r$ superimposed with the $68\%$ 
 and $95\%$ CL BICEP2/KECK Array contours in Ref.~\cite{Array:2015xqh}. 
 The black dots represent the predictions of the quadratic potential 
 $V(\phi) = m^2 \phi^2/2$, with e-folding $N_e = 50$ and $60$.  The green 
 lines are the predictions of the cosine potential, Eq.~(\ref{eq:V-natural}), 
 and the red lines are those of the (holographic) pure natural inflation 
 potential of Eq.~(\ref{eq:V_pure-nat}).  For the latter, we have varied 
 $F/M_{\rm Pl} = 0.1~\mbox{--}~100$, with $F/M_{\rm Pl} = 10, 5, 1$ 
 indicated by the red dots (from top to bottom).}
\label{fig:combined}
\end{figure}

Conventionally, the potential of the axion field as in Eq.~(\ref{eq:L-theta}) 
is assumed to take the form generated by non-perturbative instantons
\begin{equation}
  V(\phi) = \Lambda^4 \left[1 -\cos\left(\frac{\phi}{f} \right) \right],
\label{eq:V-natural}
\end{equation}
where $\Lambda$ is the dynamical scale of the Yang-Mills theory.  The resulting 
inflation model is called natural inflation~\cite{Freese:1990rb,Adams:1992bn}, 
which has been extensively studied in the literature.  The potential of 
Eq.~(\ref{eq:V-natural}), however, is not favored by the current data, and 
it would soon be excluded at a higher confidence level if the bound on $r$ 
improves with $n_s$ staying at the current value; see Fig.~\ref{fig:combined}.

It is known since long ago, however, that the cosine potential 
in Eq.~(\ref{eq:V-natural}) is not correct in general, as argued 
by Witten~\cite{Witten:1979vv,Witten:1980sp} in the large $N$ 
limit~\cite{tHooft:1973alw} with the 't~Hooft coupling $\lambda 
\equiv g^2 N$ held fixed.%
\footnote{See, e.g., Refs.~\cite{Kaloper:2011jz,Dubovsky:2011tu,%
 Dine:2014hwa,Yonekura:2014oja,Kaloper:2016fbr} for related discussion 
 in the context of inflation.}
In particular, while the physics is periodic in $\phi$ with the period 
of $2\pi f$ (because $\theta \equiv \phi/f$ is the $\theta$ angle of 
the Yang-Mills theory), the multi-valued nature of the potential allows 
for the potential of $\phi$ in a single branch
\begin{equation}
  V(\phi) = N^2 \Lambda^4\, 
    \mathcal{V} \left(\frac{\lambda \phi}{8\pi^2 N f}\right),
\label{eq:branch}
\end{equation}
{\it not} to respect the periodicity under $\phi \rightarrow \phi + 2\pi f$. 
Here, the combination
\begin{equation}
  x \equiv \frac{\lambda \phi}{8\pi^2 N f},
\label{eq:x}
\end{equation}
appearing in the argument of $\mathcal{V}(x)$ is determined by analyzing 
the large $N$ limit.  This allows for building axionic models of inflation 
in which the range of the field excursion exceeds the decay constant 
$f$~\cite{Silverstein:2008sg,McAllister:2008hb,Kaloper:2008fb}.

The potential of Eq.~(\ref{eq:branch}) has an expansion of the form
\begin{equation}
  V(\phi) = \sum_{n=1}^{\infty} b_{2n} \left(\frac{\phi}{F}\right)^{2n},
\label{eq:V-exp}
\end{equation}
where $F \propto f$.  The values of the coefficients $b_{2n}$---more precisely 
their signs and double ratios---are important for how the predictions 
for $n_s$ and $r$ change as $F$ is varied.  If the cosine potential in 
Eq.~(\ref{eq:V-natural}) were valid, then we would obtain
\begin{equation}
  \textrm{sgn}(b_{2n}) = (-1)^{n-1},
\label{eq:signs}
\end{equation}
and
\begin{equation}
  \frac{\frac{b_6}{b_4}}{\frac{b_4}{b_2}} = \frac{2}{5},
\qquad
  \frac{\frac{b_8}{b_6}}{\frac{b_6}{b_4}} = \frac{15}{28},
\qquad
  \cdots,
\label{eq:ratio-cos}
\end{equation}
which lead to the curves labeled as ``cosine'' in Fig.~\ref{fig:combined}. 
The correct values of the double ratios, however, are expected to be 
different from these values.  In fact, $b_{2n}$'s obtained by lattice 
gauge theory disfavor the cosine form of Eq.~(\ref{eq:V-natural}) 
and are rather consistent with those expected from large $N$ 
expansion~\cite{Giusti:2007tu}.

While $b_{2n}$'s may in principle be determined by lattice calculations, 
their errors are still large.  Instead, we may infer the form of the 
potential by the following arguments.  First, invariance under the $CP$ 
transformation $\phi \rightarrow -\phi$ implies that $\mathcal{V}(x)$ 
is a function of $x^2$, where we have absorbed a possible bare $\theta$ 
parameter in the definition of $\phi$.  Second, $\mathcal{V}(x)$ is expected 
to flatten as the potential energy approaches the point of the deconfining 
phase transition with increasing $|\phi|$ (since the dynamics generating 
the potential will become weaker).  Assuming that the potential is given 
by a simple power law, we thus expect $\mathcal{V}(x) \sim 1/(x^2)^p$ 
($p > 0$).  This potential is singular at $x \rightarrow 0$, and a simple 
way to regulate it is to replace $x^2$ with $x^2 + \textrm{const.}$  After 
setting the minimum of the potential to be zero, these considerations give
\begin{equation}
  V(x) = M^4 \left[ 1 - \frac{1}{\left( 1 + c x^2 \right)^p} \right] 
\quad
  (p > 0),
\label{eq:V-gen}
\end{equation}
where $M \sim \sqrt{N} \Lambda$, and $c > 0$ is a parameter of order unity. 
Here, we have used the well-established fact that the coefficient of $x^2$ 
is positive when $\mathcal{V}(x)$ is expanded around $x=0$.  We call the 
model of inflation in which the axionic inflaton potential is generated 
by a pure Yang-Mills theory (whose potential we expect to take the form 
of Eq.~(\ref{eq:V-gen})) {\it pure natural inflation}.

As in the cosine potential, the potential of Eq.~(\ref{eq:V-gen}) gives 
$\textrm{sgn}(b_{2n}) = (-1)^{n-1}$.  It, however, gives different values 
of the double ratios
\begin{equation}
  \frac{\frac{b_6}{b_4}}{\frac{b_4}{b_2}} = \frac{2(p+2)}{3(p+1)},
\;\;
  \cdots,
\;\;
  \frac{\frac{b_{2n+4}}{b_{2n+2}}}{\frac{b_{2n+2}}{b_{2n}}} 
  = \frac{(n+1)(p+n+1)}{(n+2)(p+n)},
\;\;
  \cdots.
\label{eq:ratio-gen}
\end{equation}
Therefore, predictions of this model are different from those of conventional 
natural inflation.  (For example, by equating $(b_6/b_4)/(b_4/b_2)$ we obtain 
$p = -7/2 < 0$.)  Here, we have assumed that the effect of a transition 
between different branches can be neglected, which we will argue to be 
the case.

\begin{figure}
  \includegraphics[scale=0.68]{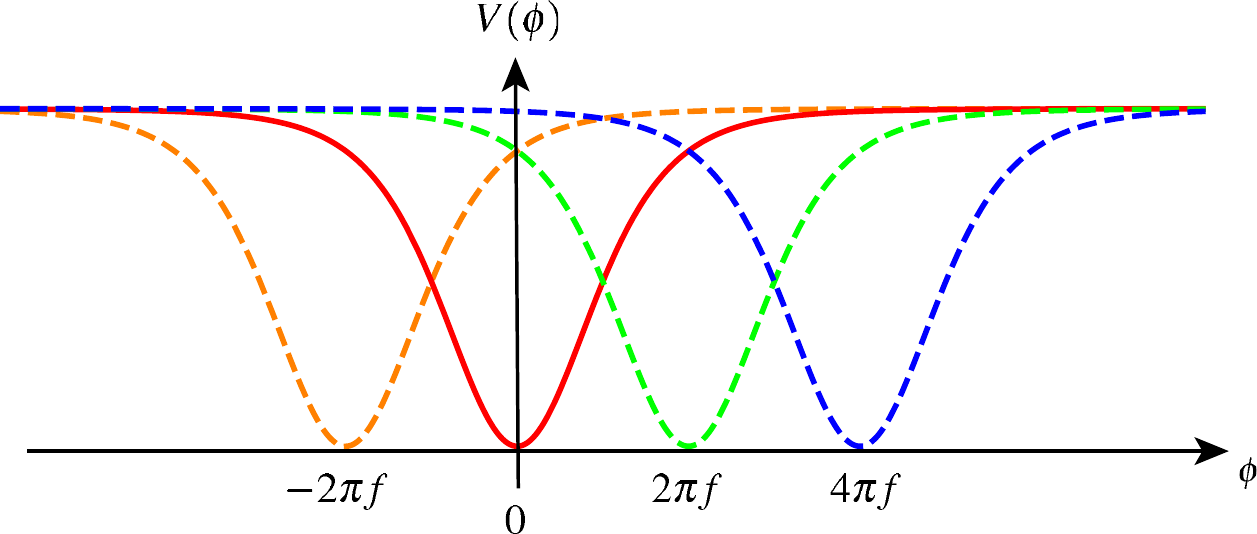}
\caption{The potential of pure natural inflation (in the holographic limit 
 $p=3$); Eq.~(\ref{eq:V_pure-nat}).  The potentials for other branches, which 
 ensure the periodicity of physics under $\phi \rightarrow \phi + 2\pi f$, 
 are also depicted by dashed lines.}
\label{fig:potential}
\end{figure}

The potential of Eq.~(\ref{eq:V-gen}) can be obtained by a holographic 
calculation~\cite{Dubovsky:2011tu,Witten:1998uka}, which is applicable in 
the limit of large $N$ and 't~Hooft coupling.  In this calculation, $N$ 
D4-branes in type~IIA string theory are considered, with the D4-branes 
wrapping a circle.  Below the Kaluza-Klein scale $M_{\rm KK}$ for the 
circle, the theory reduces to a 4d (non-supersymmetric) pure $SU(N)$ 
Yang-Mills theory, with the dynamical scale
\begin{equation}
  \Lambda = M_{\rm KK}\, e^{-\frac{24 \pi^2}{11 \lambda}},
\label{eq:Lambda-KK}
\end{equation}
where $\lambda$ is the 't~Hooft coupling at $M_{\rm KK}$.  Considering 
the backreaction to the geometry of the constant Wilson line of the 
Ramond-Ramond one-form, which represents the $\theta$ angle of the gauge 
theory, the potential of the form of Eq.~(\ref{eq:V-gen}) is obtained 
with $c=1$ and $p=3$.  Specifically, the potential of $\phi$ for a single 
branch is given by
\begin{equation}
  V(\phi) = M^4 \left[ 1 
    - \frac{1}{\Bigl( 1 + \bigl(\frac{\phi}{F}\bigr)^2 \Bigr)^3} \right],
\label{eq:V_pure-nat}
\end{equation}
where%
\footnote{The 't~Hooft coupling (and the gauge coupling squared) defined 
 in Ref.~\cite{Dubovsky:2011tu} is a factor of 2 smaller than $\lambda$ 
 ($g^2$) here.}
\begin{equation}
  M^4 = \frac{\lambda N^2}{3^7 \pi^2} M_{\rm KK}^4,
\qquad 
  F = \frac{8\pi^2 N}{\lambda} f.
\label{eq:M-F}
\end{equation}
The potentials for the other branches are obtained by replacing $\phi$ with 
$\phi + 2\pi k f$ ($k \in \mathbb{Z}$); see Fig.~\ref{fig:potential}.

To illustrate the parameter region we consider, let us choose
\begin{equation}
  \lambda \simeq 8\pi^2.
\label{eq:lambda}
\end{equation}
Strictly speaking, the holographic calculation is not quite valid with this 
value of the 't~Hooft coupling---it requires a larger value of $\lambda$. 
However, we may expect, e.g.\ based on the success of the AdS/QCD 
program~\cite{Erlich:2005qh,DaRold:2005mxj}, that this reasonably 
approximates the true dynamics of the 4d Yang-Mills theory.  With this 
choice of $\lambda$, we find
\begin{equation}
  M \approx \sqrt{N} \Lambda,
\qquad
  F \approx N f,
\label{eq:approx}
\end{equation}
as one naively expects from dimensional and $N$-scaling considerations. 
From the analysis of Ref.~\cite{Dubovsky:2011tu}, we expect that for 
sufficiently large $N$ the effect of a transition between different 
branches is not important, unless $\phi$ becomes much larger than $F$ 
(which does not occur in our analysis below).  To reproduce the observed 
amplitude of the scalar perturbation, we will need to take
\begin{equation}
  M \sim 10^{16}~{\rm GeV}.
\label{eq:M-Lambda}
\end{equation}
The precise value depends on other parameters, e.g.\ $F$.

The second expression in Eq.~(\ref{eq:approx}) implies that for $N \gg 1$ 
the characteristic scale for the field excursion, $F$, can be much larger 
than the axion decay constant $f$.  This, however, does not mean that $F$ 
can be much larger than the Planck scale $M_{\rm Pl} \simeq 1.22 \times 
10^{19}~{\rm GeV}$.  In general, the decay constant $f$ is expected to be 
smaller than the field theoretic cutoff (string) scale $M_*$
\begin{equation}
  f \lesssim M_*.
\label{eq:f-M_*}
\end{equation}
On the other hand, the Planck scale is related with $M_*$ as $M_{\rm Pl}^2 
\sim N^2 M_*^2$ (see, e.g., Ref.~\cite{Dvali:2007hz}), so that $N$ drops from 
the relation between $F$ and $M_{\rm Pl}$:
\begin{equation}
  F \lesssim O(M_{\rm Pl}).
\label{eq:F-M_Pl}
\end{equation}
We argue that this is a desired feature.  If $F \gg M_{\rm Pl}$, the 
inflaton potential would be well approximated by the first, quadratic 
term in Eq.~(\ref{eq:V-exp}), which is excluded by the data.  Because 
of Eq.~(\ref{eq:F-M_Pl}), however, we expect that higher terms in 
Eq.~(\ref{eq:V-exp}) are important.  This makes the predictions of 
the model deviate from those of the quadratic potential $V(\phi) = 
m^2 \phi^2/2$.

In Fig.~\ref{fig:combined}, we plot the values of $n_s$ and $r$ predicted 
by the potential of Eq.~(\ref{eq:V_pure-nat}).  For $F \gg M_{\rm Pl}$, 
the predictions approach those of the quadratic potential, denoted by the 
black dots for $e$-folding $N_e = 50$ (left) and $60$ (right).  As $F$ 
decreases, however, they deviate from these values.  In particular, the 
tensor-to-scalar ratio $r$ decreases while $n_s$ being (almost) kept, as 
represented by the lines indicated as ``pure natural.''  This is because 
higher terms in Eq.~(\ref{eq:V-exp}) start contributing.%
\footnote{A special case of $V(\phi) \approx A - B/\phi^6$ leading to 
 $n_s \simeq 0.965$ and $r \simeq 8 \times 10^{-4}$ was discussed in 
 Ref.~\cite{Dubovsky:2011tu}.}
In the figure, we have varied $F$ from $100 M_{\rm Pl}$ to $0.1 M_{\rm Pl}$, 
with the predictions for $F/M_{\rm Pl} = 10, 5, 1$ indicated by the red 
dots (from top to bottom).  We find that the model gives the values of 
$n_s$ and $r$ consistent with the data at the $95\%$ ($68\%$) CL for 
$F/M_{\rm Pl} \lesssim 3.3$ ($0.7$) for $N_e = 50$ and $F/M_{\rm Pl} 
\lesssim 6.8$ ($4.4$) for $N_e = 60$.  Given Eq.~(\ref{eq:F-M_Pl}), this 
is quite satisfactory.

\begin{figure}
  \includegraphics[scale=0.68]{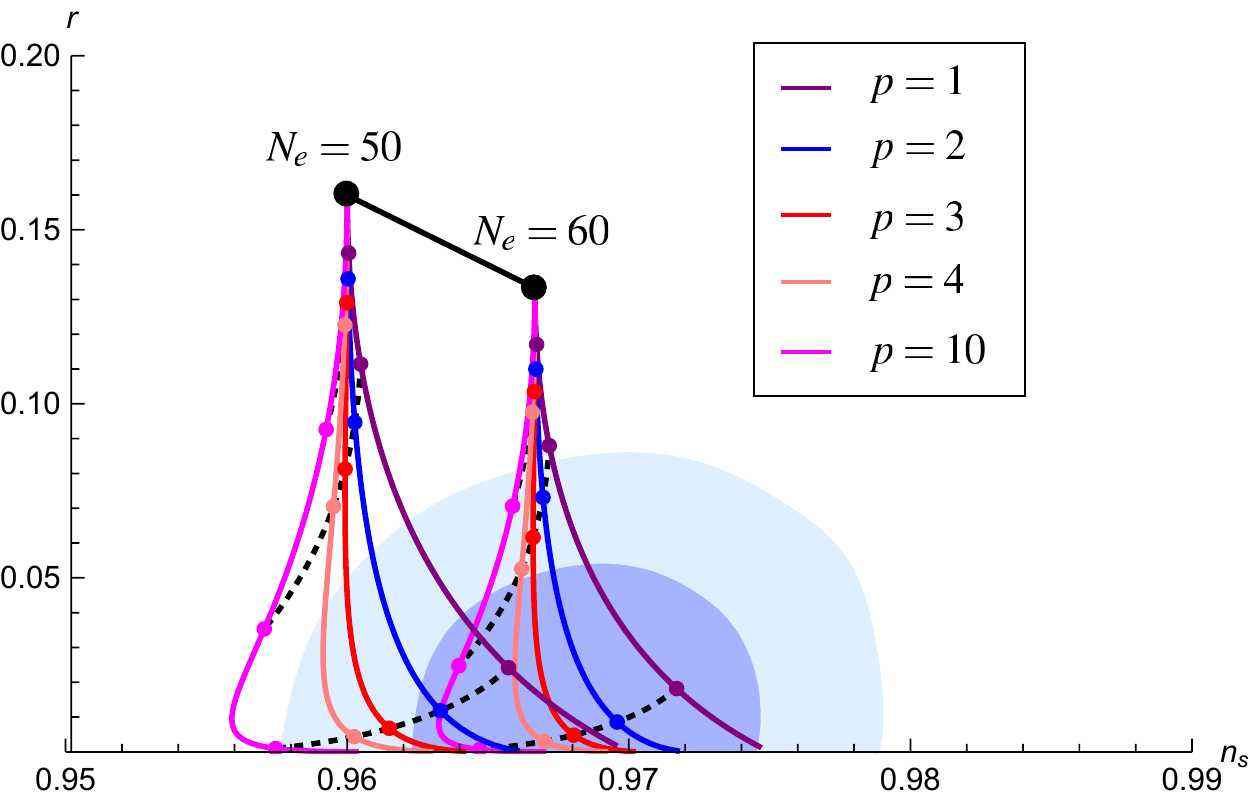}
\caption{Same as in Fig.~\ref{fig:combined} but for pure natural 
 inflation potentials with various values of $p = 1, 2, 3, 4, 10$; see 
 Eq.~(\ref{eq:V_pure-nat-gen}).}
\label{fig:p-varied}
\end{figure}
In Fig.~\ref{fig:p-varied}, we plot the predictions arising from the potential 
in which $p$ in Eq.~(\ref{eq:V-gen}) takes more general values $p = 1, 2, 3, 
4, 10$.  We have parameterized the potentials as
\begin{equation}
  V(\phi) = M^4 \left[ 1 
    - \frac{1}{\Bigl( 1 + \bigl(\frac{\phi}{F}\bigr)^2 \Bigr)^p} \right],
\label{eq:V_pure-nat-gen}
\end{equation}
and, as in Fig.~\ref{fig:combined}, varied $F/M_{\rm Pl}$ in the range 
between $100$ and $0.1$, with the dots representing $F/M_{\rm Pl} = 
10, 5, 1$ (from top to bottom).  We find that the success of the model is 
robust for a wide range of $p \approx 1~\mbox{--}~O(10)$:\ the predicted 
values of $n_s$ and $r$ agree well with the current data for $F/M_{\rm Pl} 
\lesssim O(1)$.  We thus conclude that pure natural inflation is consistent 
with the data even if the true potential does not take exactly the form of 
Eq.~(\ref{eq:V_pure-nat}) as suggested by the holographic analysis.

So far, we have focused on the case that the gauge group of the Yang-Mills 
theory is $SU(N)$.  However, our basic arguments, e.g.\ those around 
Eq.~(\ref{eq:V-gen}), do not depend on this specific choice.  We thus 
expect that similar predictions also result for other gauge groups, 
with $N$ replaced by the dual Coxeter number of the group.

Finally, we mention that a large value of $N$---despite the fact that it 
does not help to make $F$ larger than $M_{\rm Pl}$---can make the decay 
constant $f$ smaller than $F$; see Eq.~(\ref{eq:approx}).  This allows 
for enhancing couplings of the inflaton $\phi$ to the standard model 
gauge fields for fixed $F$, i.e.\ for a fixed inflation potential.  This 
in turn allows for raising reheating temperature $T_R$.  The reheating 
temperature is given by $T_R \simeq 0.2 \sqrt{\Gamma_\phi M_{\rm Pl}}$, 
where
\begin{equation}
  \Gamma_\phi \approx \frac{n m_\phi^3}{4096 \pi^5 f^2},
\label{eq:inf-decay}
\end{equation}
is the inflaton decay width.  Here, 
\begin{equation}
  m_\phi = \sqrt{2p} \frac{M^2}{F} 
  \simeq \frac{10^{13}~{\rm GeV}}{F/M_{\rm Pl}},
\label{eq:inf-mass}
\end{equation}
is the inflaton mass, and $n$ is the final state multiplicity.  ($n = 12$ 
if $\phi$ decays only to the standard model gauge fields.)  This yields
\begin{equation}
  T_R \sim 10^9~{\rm GeV} \left( \frac{N}{10} \right) 
    \left( \frac{0.5}{F/M_{\rm Pl}} \right)^{5/2},
\label{eq:T_R}
\end{equation}
The model can, therefore, be made consistent with thermal leptogenesis, which 
requires $T_R \gtrsim 2 \times 10^9~{\rm GeV}$~\cite{Buchmuller:2005eh}, 
even for $F/M_{\rm Pl} \approx O(1)$.

\vspace{0.3cm}

\begin{acknowledgments}
This work was supported in part by the WPI Research Center Initiative (MEXT, 
Japan).  The work of Y.N. was also supported in part by the National Science 
Foundation under grants PHY-1521446, by MEXT KAKENHI Grant Number 15H05895, 
and by the Department of Energy (DOE), Office of Science, Office of High Energy 
Physics, under contract No.\ DE-AC02-05CH11231.  The work of M.Y. was supported 
by JSPS KAKENHI (15K17634) and JSPS-NRF Joint Research Project.
\end{acknowledgments}


\begin{thebibliography}{99}

\bibitem{Ade:2015lrj}
P.~A.~R.~Ade {\it et al.} [Planck Collaboration],
``Planck 2015 results. XX. Constraints on inflation,''
Astron.\ Astrophys.\  {\bf 594}, A20 (2016)
[arXiv:1502.02114 [astro-ph.CO]].

\bibitem{Array:2015xqh} 
P.~A.~R.~Ade {\it et al.} [BICEP2 and Keck Array Collaborations],
``Improved constraints on cosmology and foregrounds from BICEP2 and Keck Array cosmic microwave background data with inclusion of 95 GHz band,''
Phys.\ Rev.\ Lett.\  {\bf 116}, 031302 (2016)
[arXiv:1510.09217 [astro-ph.CO]].

\bibitem{Linde:1983gd}
A.~D.~Linde,
``Chaotic inflation,''
Phys.\ Lett.\  {\bf 129B}, 177 (1983).

\bibitem{Freese:1990rb}
K.~Freese, J.~A.~Frieman and A.~V.~Olinto,
``Natural inflation with pseudo Nambu-Goldstone bosons,''
Phys.\ Rev.\ Lett.\  {\bf 65}, 3233 (1990).

\bibitem{Adams:1992bn}
F.~C.~Adams, J.~R.~Bond, K.~Freese, J.~A.~Frieman and A.~V.~Olinto,
``Natural inflation:\ particle physics models, power law spectra for large scale structure, and constraints from COBE,''
Phys.\ Rev.\ D {\bf 47}, 426 (1993)
[hep-ph/9207245].

\bibitem{Witten:1979vv}
E.~Witten,
``Current algebra theorems for the U(1) Goldstone boson,''
Nucl.\ Phys.\ B {\bf 156}, 269 (1979).

\bibitem{Witten:1980sp}
E.~Witten,
``Large N chiral dynamics,''
Annals Phys.\ {\bf 128}, 363 (1980).

\bibitem{tHooft:1973alw}
G.~'t Hooft,
``A planar diagram theory for strong interactions,''
Nucl.\ Phys.\ B {\bf 72}, 461 (1974).

\bibitem{Kaloper:2011jz}
N.~Kaloper, A.~Lawrence and L.~Sorbo,
``An ignoble approach to large field inflation,''
JCAP {\bf 03}, 023 (2011)
[arXiv:1101.0026 [hep-th]].

\bibitem{Dubovsky:2011tu}
S.~Dubovsky, A.~Lawrence and M.~M.~Roberts,
``Axion monodromy in a model of holographic gluodynamics,''
JHEP {\bf 02}, 053 (2012)
[arXiv:1105.3740 [hep-th]].

\bibitem{Dine:2014hwa}
M.~Dine, P.~Draper and A.~Monteux,
``Monodromy inflation in SUSY QCD,''
JHEP {\bf 07}, 146 (2014)
[arXiv:1405.0068 [hep-th]].

\bibitem{Yonekura:2014oja}
K.~Yonekura,
``Notes on natural inflation,''
JCAP {\bf 10}, 054 (2014)
[arXiv:1405.0734 [hep-th]].

\bibitem{Kaloper:2016fbr}
N.~Kaloper and A.~Lawrence,
``London equation for monodromy inflation,''
Phys.\ Rev.\ D {\bf 95}, 063526 (2017)
[arXiv:1607.06105 [hep-th]].

\bibitem{Silverstein:2008sg}
E.~Silverstein and A.~Westphal,
``Monodromy in the CMB:\ gravity waves and string inflation,''
Phys.\ Rev.\ D {\bf 78}, 106003 (2008)
[arXiv:0803.3085 [hep-th]].

\bibitem{McAllister:2008hb}
L.~McAllister, E.~Silverstein and A.~Westphal,
``Gravity waves and linear inflation from axion monodromy,''
Phys.\ Rev.\ D {\bf 82}, 046003 (2010)
[arXiv:0808.0706 [hep-th]].

\bibitem{Kaloper:2008fb}
N.~Kaloper and L.~Sorbo,
``A natural framework for chaotic inflation,''
Phys.\ Rev.\ Lett.\ {\bf 102}, 121301 (2009)
[arXiv:0811.1989 [hep-th]].

\bibitem{Giusti:2007tu}
L.~Giusti, S.~Petrarca and B.~Taglienti,
``Theta dependence of the vacuum energy in the SU(3) gauge theory from the lattice,''
Phys.\ Rev.\ D {\bf 76}, 094510 (2007)
[arXiv:0705.2352 [hep-th]].

\bibitem{Witten:1998uka}
E.~Witten,
``Theta dependence in the large N limit of four-dimensional gauge theories,''
Phys.\ Rev.\ Lett.\ {\bf 81}, 2862 (1998)
[hep-th/9807109].

\bibitem{Erlich:2005qh}
J.~Erlich, E.~Katz, D.~T.~Son and M.~A.~Stephanov,
``QCD and a holographic model of hadrons,''
Phys.\ Rev.\ Lett.\ {\bf 95}, 261602 (2005)
[hep-ph/0501128].

\bibitem{DaRold:2005mxj}
L.~Da Rold and A.~Pomarol,
``Chiral symmetry breaking from five dimensional spaces,''
Nucl.\ Phys.\ B {\bf 721}, 79 (2005)
[hep-ph/0501218].

\bibitem{Dvali:2007hz}
G.~Dvali,
``Black holes and large N species solution to the hierarchy problem,''
Fortsch.\ Phys.\ {\bf 58}, 528 (2010)
[arXiv:0706.2050 [hep-th]].

\bibitem{Buchmuller:2005eh}
W.~Buchm\"{u}ller, R.~D.~Peccei and T.~Yanagida,
``Leptogenesis as the origin of matter,''
Ann.\ Rev.\ Nucl.\ Part.\ Sci.\  {\bf 55}, 311 (2005)
[hep-ph/0502169].

\end{thebibliography}
\end{document}